\documentclass[useAMS,usenatbib,usegraphicx]{mn2e}
\usepackage{amsmath}

\title[Very High Energy $\gamma$-ray emission from RBS 0970]{Very High Energy $\gamma$-ray emission from RBS 0970}

\author[Anthony M. Brown]{Anthony M. Brown$^{1}$\thanks{E-mail: anthony.brown@durham.ac.uk}\\
$^{1}$Department of Physics and Astronomy, University of Durham, South Road, Durham, DH1 3LE, UK}
\begin{document}

\date{Accepted 2014 April 13. Received 2014 April 01. In original form 2014 January 31}

\pagerange{\pageref{firstpage}--\pageref{lastpage}} \pubyear{2014}

\maketitle

\label{firstpage}

\begin{abstract}
In this letter I report the \textit{Fermi} Large Area Telescope (LAT) detection of Very High Energy (VHE; $E>100$ GeV) $\gamma$-ray emission from the BL Lac object RBS 0970. 5.3 years of LAT observations revealed the presence of 3 VHE photon events within 0\ensuremath{^{\circ}}.1 of RBS 0970, with a subsequent unbinned likelihood analysis finding RBS 0970 to be a source of VHE photons at the $6.5\sigma$ level of confidence. The $\geq1$ GeV flux, binned in monthly periods, did not indicate any flux brightening of RBS 0970 accompanying the emission of the VHE photons. However, a likelihood analysis of the $0.1-100$ GeV flux, binned in 28 day periods centered on detection of the VHE photons, revealed that the emission of the lowest energy VHE photons coincided with a hardening of the $\gamma$-ray spectrum. Interestingly, the same analysis did not find any significant $\gamma$-ray emission from RBS 0970 during the emission of the highest energy VHE event. The discovery of RBS 0970 as a VHE emitter, combined with the spectral variability, suggest RBS 0970 to be a good candidate for follow-up observations with ground-based $\gamma$-ray observatories.
\end{abstract}


\section{INTRODUCTION}
The \textit{Fermi} $\gamma$-ray Space Telescope affords an ideal opportunity to investigate the inner workings of Active Galactic Nuclei (AGN). Since 2008 August 4, the vast majority of data taken by \textit{Fermi} has been in the default \textit{all-sky-survey} observing mode, whereby the Large Area Telescope (LAT) onboard \textit{Fermi} points away from the Earth and rocks north and south of its orbital plane. This rocking motion, coupled with \textit{Fermi}-LAT's large effective area, allows \textit{Fermi} to scan the entire $\gamma$-ray sky every two orbits, or approximately every three hours (\citet{ritz}). With this ability to scan the sky every 3 hours, the LAT has allowed us to catch AGN during brief flares of $\gamma$-ray activity (e.g. \citet{dickinson}), with these flares sometimes resulting in the discovery of Very High Energy (VHE; $E_{\gamma}>100$ GeV) emission from the flaring AGN (e.g. \citet{ong} \& \citet{aliu}). 

While it's 3 hour scan period is important for catching brief periods of flare activity from AGN, coupling \textit{Fermi}-LAT's continual scanning of the sky with a long mission livetime allows us to construct a deep exposure of the extragalactic sky. This deep exposure affords us the ability to perform searches for faint VHE sources which would otherwise go undetected by the pointed observations of  ground-based Imaging Atmospheric Cherenkov Telescope (IACT) arrays.

RBS 0970 is a point-like radio source, with a redshift of z$=0.124$ (\citet{gommi}). Detected by successive X-ray surveys using the EINSTEIN, ROSAT and BEPPO-SAX satellites (\citet{einstein}, \citet{rosat},  \citet{beppo}), RBS 0970 has been optically identified as a BL Lac object with SDSS observations (\citet{sdss}). As a member of the blazar sub-class of AGN, it is no surprise that RBS 0970 is present in both the 1 and 2 year \textit{Fermi}-LAT AGN catalogues (\citet{1agn}; \citet{acker3}). Furthermore, the recent $E_{\gamma}>10$ GeV LAT catalogue also lists RBS 0970 as a source (\citet{hecat}).

This letter reports the discovery of VHE emission from RBS 0970. Utilising 5.3 years of \textit{Fermi}-LAT data, 3 \textsc{ultraclean} events were discovered to be clustered within 0\ensuremath{^{\circ}}.1 of RBS 0970. The emission of some of these VHE photons was observed to occur during periods of spectral hardening, suggesting a harder-when-brighter property for the VHE emission. In \textsection 2 the \textit{Fermi}-LAT observations and analysis routines used in this study are described, along with the results of the 1-300 GeV likelihood analysis. The results on the VHE emission study of RBS 0970 are shown in \textsection 3. A brief investigation into the global $\gamma$-ray characteristics of RBS 0970 when the VHE emission occurs is presented in \textsection 4, with the conclusions given in \textsection 5.

\section{\textit{Fermi}-LAT OBSERVATIONS AND DATA ANALYSIS}

The data used in this study comprises all \textit{Fermi}-LAT event and spacecraft data taken during the first 5.3 years of \textit{Fermi}-LAT operation, from 2008 August 4 to 2013 December 12, which equates to a Mission Elapse Time (MET) interval of 239557417 to 408871812. All \textsc{clean} $\gamma$-ray events\footnote{\textsc{clean} events have an event class of 3 in the \textsc{pass}7\_\textsc{rep} data.}, in the $1 < E_{\gamma} < 300$ GeV energy range, within a 5\ensuremath{^{\circ}} radius of interest (RoI) centered on the Second Fermi Source Catalogue (2FGL; \cite{nolan}) position of RBS 0970, ($\alpha_{J2000}$, $\delta_{J2000}=170$\ensuremath{^{\circ}}.2, $42$\ensuremath{^{\circ}}.2039), were considered. In accordance with the \textsc{pass}7\_\textsc{rep} criteria, a zenith cut of 100\ensuremath{^{\circ}} was applied to the data to remove any cosmic ray induced $\gamma$-rays from the limb of the Earth's atmosphere. The good time intervals were generated by applying a filter expression of ``\textsc{(data\_qual$==$1) \&\& (lat\_config$==$1) \&\& abs(rock\_angle)$<$ 52}'' to the data. 

Throughout this analysis, version \textsc{v9r32p5} of the \textit{Fermi Science Tools} was used in conjunction with the  \textsc{p7rep\_clean\_v15} instrument response functions (IRFs). During the analysis, a model file consisting of both point and diffuse sources of $\gamma$-rays was utilised. In particular, the model file consisted of the most recent Galactic, gll\_iem\_v05.fit, and extragalactic, iso\_clean\_v05.txt, diffuse models, and all $\gamma$-ray point sources within a 6\ensuremath{^{\circ}} RoI centered on RBS 0970. The positions of these point sources, along with their spectral shapes, were taken from the 2FGL. The normalisation factor of the extragalactic diffuse emission was left free to vary, while the Galactic diffuse template was multiplied by a power law in energy, the normalisation of which was left free to vary.

Firstly a binned maximum likelihood analysis was performed on the entire 5.3 year data set. For RBS 0970 itself, a power law spectral shape of the form $dN/dE = $ A$ \times (E/E_o)^{-\Gamma}$ was assumed, with the normalisation, A, and the spectral index, $\Gamma$, left free to vary\footnote{It should be noted that the likelihood analysis was also applied to the 5.3 year data set assuming both a broken power law and log parabola description of RBS 0970's spectrum; however, the power law was found to have the highest significance.}. The normalisation and spectral parameters of all point sources within 5\ensuremath{^{\circ}} of RBS 0970 were left free to vary, while the normalisation and spectral parameters for all point sources within an annulus of 5\ensuremath{^{\circ}} to 6\ensuremath{^{\circ}} from RBS 0970 were fixed to those published in the 2FGL.  

Utilising the above described model, the binned likelihood analysis of the 5.3 year data set resulted in the following best-fit power law function for RBS 0970:

\begin{equation}
 \dfrac{dN}{dE}= (2.3 \pm 0.1) \times 10^{-13} (\dfrac{E}{3795.9\text{ MeV}})^{-2.25\pm0.06} \nonumber
\end{equation}

\begin{equation}
 \text{ photons cm}^{-2} \text{s}^{-1} \text{MeV}^{-1}
\end{equation}

which equates to an integrated flux, in the $1-300$ GeV energy range, of

\begin{equation}
  F_{E>1\text{ GeV}} = (3.69 \pm 0.18) \times 10^{-9}  \text{ photons cm}^{-2} \text{s}^{-1}
\end{equation}

taking only statistical errors into account\footnote{Primarily governed by the uncertainty in the effective area, the systematic uncertainty of the integrated flux is energy dependent and is currently estimated as 10\% at 100 MeV, down to 5\% at 560 MeV and back to 10\% for 10 GeV photons (\citet{acker1}).}. For the best-fit power law description, a test statistic\footnote{The test statistic, TS, is defined as twice the difference between the log-likelihood of two different models, $TS=2[\text{log} L - \text{log} L_{0}]$, where $L$ and $L_{0}$ are defined as the likelihood when the source is included or not respectively (\citet{mattox2}).} of $TS=2132.6$ was found, corresponding to a $\sim46\sigma$ detection of RBS 0970 above 1 GeV. 

\begin{figure}
 \centering
\includegraphics[width=80mm]{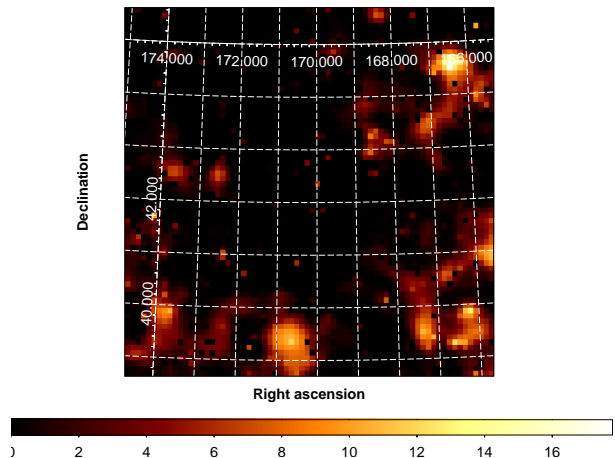}
\caption{A TS map of the $1-300$ GeV events during the entire 5.3 year period, centered on the co-ordinates of RBS 0970. The colour scale is the TS value of the individual pixels, each of which are $0.1 \times 0.1$ degrees. The largest excess in the TS map is $TS\simeq16$, which implies that there are no significant sources in the data that are not present in the model. As such, the model used to describe the data is an accurate description of the data recorded.}
\label{tsmap}
\end{figure}

\begin{table*}
 \begin{minipage}{150mm}
   \caption{Summary of the three VHE events from RBS 0970 detected by \textit{Fermi}-LAT. It should be noted that all 3 of these events are also \textsc{ultraclean} class events.}
   \begin{center}
     \begin{tabular}{cccccc} \hline \hline
      Energy & MET  & MJD     & $\alpha_{J2000}$ & $\delta_{J2000}$ &  \textsc{gtsrcprob} \\ 
      (GeV)    &   (second) & (day)         & (deg)        & (deg)         & probability \\ \hline
      114    & 318444536.396 &   55595.07014 & 170.119      & 42.264       & 0.9987601   \\ 
           &    &        &      &              &             \\
      273    & 354348414.083 &  56011.25553 & 170.225      & 42.189       & 0.9999382   \\ 
           &    &        &      &              &             \\
      117    & 376544517.873  & 56268.15488 & 170.168      & 42.283       & 0.9986324    \\   \hline \hline
    \end{tabular}
  \end{center}
  \label{photondetails}
\end{minipage}
\end{table*}

The long exposure of the 5.3 year integration can result in additional faint sources being present in the data that were not present in the 2FGL. If these sources are present in the data, and not properly accounted for within the model file, they can artifically increase the significance of the $\gamma$-ray flux from RBS 0970 (e.g. \citet{mepicA} \& \citet{oscar}). To check if indeed any additional sources were present, the best-fit model of the $1-300$ GeV events, in conjunction with the \textsc{gttsmap} \textit{Fermi} tool, was used to construct a TS significance map of the $1-300$ GeV events that were observed during the 5.3 year period. The resultant map, centered on RBS 0970, can be seen in Figure \ref{tsmap}. The largest excess hotspot in the TS map is $TS\simeq16$, located $\sim3.7$\ensuremath{^{\circ}} from RBS 0970. As such, while the moderate size of the TS hotspot possibly hints at the presence of a new source, the angular separation between the hotspot and RBS 0970 minimises any impact it has on the likelihood fit of RBS 0970\footnote{Below 10 GeV photon energy, the 68\% containment angle of the photon direction is approximately given by $\theta \simeq 0.8$\ensuremath{^{\circ}}($E_\gamma/$GeV)$^{-0.8}$, with the 95\% containment angle being less than 1.6 times the angle for 68\% containment. As such, 1.3\ensuremath{^{\circ}} is the 95\% containment angle for a 1 GeV photon.}, in the $1-300$ GeV energy range.

\section{Very High Energy $\gamma$-ray properties}

A closer inspection of the individual photon events within 0\ensuremath{^{\circ}}.1 of RBS 0970 revealed the presence of 3 VHE $\gamma$-rays. All three VHE events are also classed as \textsc{ultraclean} events, a subclass of \textsc{clean} events that have the highest probability of being photons. Utilising the combined diffuse and point source model file, with all normalisation and spectral parameters frozen to the best-fit values of the 5.3 year binned likelihood analysis, the \textsc{gtsrcprob} \textit{Fermi} tool was used to calculate the probability that each of the VHE $\gamma$-ray events originated from RBS 0970, as opposed to other sources such as the Galactic or extragalactic diffuse emission. The results of these probability calculations are shown in Table \ref{photondetails}, along with the energy, time and ($\alpha_{J2000}$, $\delta_{J2000}$) of each VHE photon. 

Considered in isolation, none of these $E_{\gamma}>100$ GeV events are significant enough to consider RBS 0970 a source of VHE $\gamma$-rays; the most significant event being the 273 GeV photon detected on MJD$=56011.25553$, with a $\sim4\sigma$ significance of originating from RBS 0970. The other two VHE events have a $\sim3\sigma$ confidence of originating from RBS 0970. 
 
While these 3 VHE photon events are not signicant when considered individually, the clustering of such energetic photons within a relatively small area can be significant given the small background rates detected by the \textit{Fermi}-LAT above 100 GeV (e.g. \citet{neronov}, \citet{neronov2} \& \citet{tanaka}). To determine if RBS 0970 is indeed a source of VHE $\gamma$-ray photons, an unbinned likelihood analysis was applied to all $E_{\gamma}>100$ GeV \textsc{clean} events within 5\ensuremath{^{\circ}} of RBS 0970 for the entire 5.3 year data set. The likelihood analysis was applied utilising the combined diffuse and point source model from \textsection 2, with the normalisation and spectral parameters left free to vary. The resultant best-fit power law function for RBS 0970, in the $100-300$ GeV energy range, was found to be:

\begin{equation}
 \dfrac{dN}{dE}= (0.9 \pm 5.2) \times 10^{-13} (\dfrac{E}{3795.9\text{ MeV}})^{-1.71\pm1.56} \nonumber
\end{equation}

\begin{equation}
 \text{ photons cm}^{-2} \text{s}^{-1} \text{MeV}^{-1}
\end{equation}

which equates to an integrated flux of

\begin{equation}
  F_{E>100\text{ GeV}} = (2.47 \pm 1.26) \times 10^{-11}  \text{ photons cm}^{-2} \text{s}^{-1}
\end{equation}

again taking only statistical errors into account. The TS value of the best-fit power law was $TS=41.9$, equating to a significance of $\sim6.5\sigma$. As such, this analysis represents the discovery of RBS 0970 as a source of VHE $\gamma$-rays. 

To localise the origin of the VHE $\gamma$-ray emission, another \textit{Fermi} tool, \textsc{gtfindsrc}, was applied to all $E_{\gamma}>100$ GeV events within 5\ensuremath{^{\circ}} of RBS 0970. Using the same combined diffuse and point source model that was applied during the \textsc{gtsrcprob} routine, the observed VHE $\gamma$-ray emission was found to originate from the point ($\alpha_{J2000}$, $\delta_{J2000}=$ 170\ensuremath{^{\circ}}.169, 42\ensuremath{^{\circ}}.2476), with a 95\% error radius of 0\ensuremath{^{\circ}}.096. As such, the VHE emission is found to be spatially co-incident with the 2FGL position of RBS 0970. 

\section{DISCUSSION}

\begin{figure*}
 \centering
 \begin{minipage}{160mm}
\includegraphics[width=180mm]{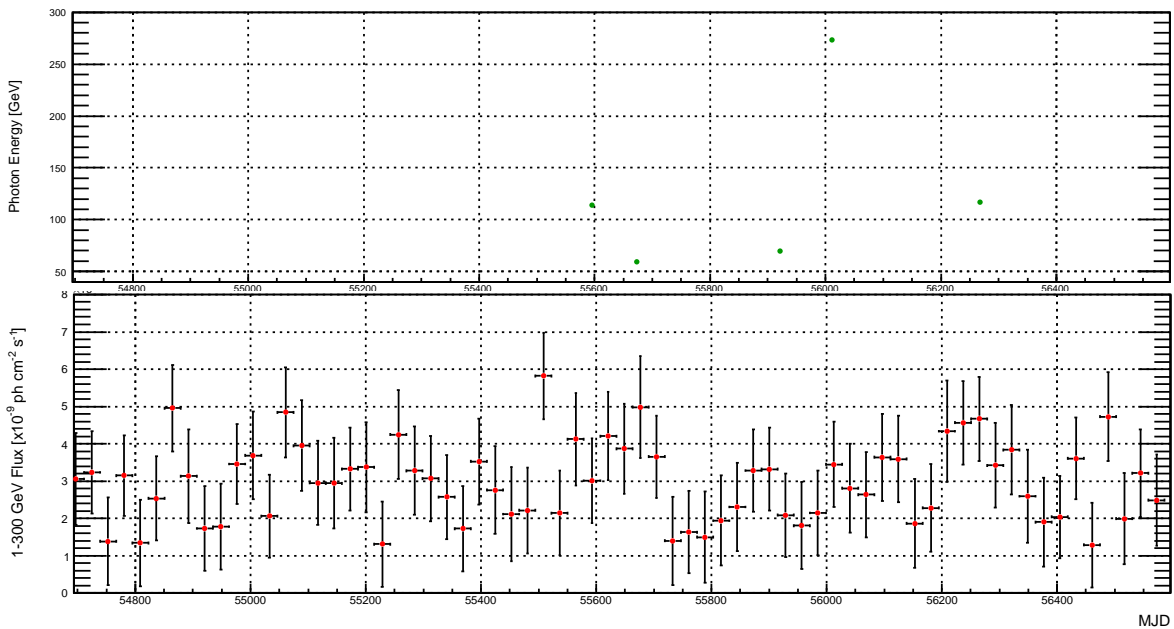}
\caption{\textit{Top panel}: Arrival times in MJD of all $E_{\gamma}>50$ GeV \textsc{clean} events within 0\ensuremath{^{\circ}}.1 of RBS 0970. All $E_{\gamma}>50$ GeV events occurred within a $\sim1.8$ year period starting with the first detected VHE photon on 55595.07014. The energy resolution for events $E_{\gamma}>10$ GeV is $\sim10$\% (\citet{acker1}). \textit{Bottom panel}: 28 day-binned lightcurve of $E_{\gamma}>1$ GeV flux from RBS 0970. There does not appear to be any marked difference in the 28 day-binned $E_{\gamma}>1$ GeV flux during the 1.8 year period when the VHE emission was observed compared to the other times of the 5.3 year period.}
\label{lc}
\end{minipage}
\end{figure*}

\citet{kat} and \citet{mengc} found that, for NGC 1275, it is the $\geq1$ GeV $\gamma$-ray flux and spectral shape that are important when triggering ground-based VHE $\gamma$-ray observations, with a higher $\geq1$ GeV flux, or harder $\gamma$-ray spectrum, more likely to be associated with the emission of VHE $\gamma$-ray photons. However, \citet{mepks} found that this was not universally applicable to all VHE emitting AGN, with no such trend being observed for PKS 1510-089. To investigate if this applies to RBS 0970, the arrival times of the $\geq50$ GeV photons within 0\ensuremath{^{\circ}}.1 of RBS 0970 were compared to the $E_{\gamma}>1$ GeV flux lightcurve; the comparison can be seen in Figure \ref{lc}. The lightcurve was constructed using the aperture photometry method, with an RoI of 1.3\ensuremath{^{\circ}} centered on RBS 0970, binned in 28-day temporal bins. As can be seen in Figure \ref{lc}, there does not appear to be any difference in the $E_{\gamma}>1$ GeV flux during the 1.8 year period when the VHE emission was observed compared to the other times of the total 5.3 year observing period. 

To further investigate the global $\gamma$-ray characteristics of RBS 0970 during the emission of the 3 VHE photons, an unbinned likelihood analysis was applied to the 0.1 $< E_{\gamma} <$ 100 GeV energy range, within $\pm14$ days the VHE event's detection. The upper energy limit of 100 GeV was chosen so as to remove any possible bias to harder spectral indices caused by the presense of the VHE events. To allow for the larger containment angle of the MeV photons, a 10\ensuremath{^{\circ}} RoI was considered. This increase in RoI also required additional point sources to be included in the model file, with the position and spectral shape of these additional point sources taken from the 2FGL. Firstly a binned analysis was applied to all events in the 5.3 year data set to find the best-fit spectral values for all point sources. Once the best-fit to the 5.3 data set was found, only data in a 28 day window centered on the detection of the individual VHE events, was considered thereafter. The model file utilised in these subsequent \textsc{gtlike} fits had all the point sources spectral and normalisation parameters frozen to the best-fit values of the 5.3 year analysis, except for the normalisation and spectral index of RBS 0970, along with the normalisation of the Galactic and extra-galactic diffuse emission, which were left free to vary. 

Interestingly, within the $0.1-100$ GeV energy range, RBS 0970 was detected during the 28 day window centered on the detection of the 114 and 117 GeV photon, but not during the 28 day window centered on the detection of the 273 GeV photon. Given that RBS 0970 was detected with a large significance during this period in the $1-300$ GeV lightcurve of Figure 2, the non-detection of RBS 0970 in the $0.1-100$ GeV energy range during the detection of the 273 GeV event suggests a large amount of significance is associated with this one VHE event. Indeed, this is what is found with the \textsc{gtsrcprob} analysis in \textsection 3. Nonetheless, the non-detection of RBS 0970 in the $0.1-100$ GeV energy range during the detection of the 273 GeV event possibly suggests that processes other than the traditional Synchrotron Self-Compton\footnote{The $\gamma$-ray flux from BL Lac object subclass of AGN is often attributed to the Synchrotron Self-Compton (SSC) model, whereby the observed $\gamma$-ray flux is produced through the inverse comptonisation of synchrotron photons by a population of relativistic electrons (eg. \citet{kraw}; \citet{mephd}; \citet{abram}).} are responsible for, at least, a fraction of the observed VHE events.

The spectral indices of the power law fits for the observed $0.1-100$ GeV flux during the detection of 114 and 117 GeV photons are shown in Table \ref{indexdetails}. In the 2-year LAT AGN catalogue (2LAC; \citet{acker3}), RBS 0970 was found to have a spectral index of $1.61$, while the 5.3 year analysis reported in \textsection 2 found that the spectral index above 1 GeV softened to $2.25\pm0.06$. These results suggest some amount of spectral variability in RBS 0970's $\gamma$-ray spectrum. As can be seen in Table \ref{indexdetails}, this spectral variability appears to be a common feature during the detection of the 114 and 117 GeV VHE photon events. When compared to the best-fit spectral index for the entire 5.3 period, $\Gamma = 2.79 \pm 0.01$, there is a clear departure towards a spectral hardening during the emission of the 114 and 117 GeV photon events; $\Gamma = 1.76 \pm 0.28$ and $\Gamma = 1.35 \pm 0.32$ respectively. This correlation suggests that the VHE emission from RBS 0970 is associated with severe hardening of the $\gamma$-ray spectrum. It is worth noting that phenomenological studies by \textit{Fermi} have found the majority of TeV bright AGN have a $\Gamma < 2$ (\citet{abdo}).

\begin{table}
   \caption{Summary of spectral indices of $0.1-100$ GeV power law fit for a period $\pm14$ days from the detection of each VHE event. It should be noted that RBS 0970 was not detected in the $0.1-100$ GeV energy range during the 28 day period centered on the detection of the 273 GeV event.}
   \begin{center}
     \begin{tabular}{ccccc} \hline \hline
      Energy & T$_{start}$ & T$_{stop }$ & $\Gamma$ & $\Delta \Gamma$ \\  \hline
      114    & 55581.70139 &  55609.70139 & 1.76  & 0.28              \\ 
      117    & 56254.15487 &  56282.15487 & 1.35  & 0.32              \\   \hline \hline
    \end{tabular}
  \end{center}
  \label{indexdetails}
\end{table}

The detection of VHE photons and the hardness of the spectrum suggest RBS 0970 is a promising target for follow-up observations with IACT arrays such as VERITAS or the future CTA (\citet{veritas} \& \citet{cta}). Such observations would allow us not only to confirm the discovery of RBS 0970 as a source of VHE photons, but also to see if the VHE emission from RBS 0970 is indeed brighter when the spectrum is harder. Moreover, ground-based observations are necessary to investigate whether the absence of significant $0.1-100$ GeV flux during the emission of the 273 GeV photon is a due to emission mechanisms other than the SSC model, or simply an artifact of limited photon statistics detected by the \textit{Fermi}-LAT.  

Nonetheless, with the detection of 3 \textsc{ultraclean} events within 0.1\ensuremath{^{\circ}}, the VHE detection of RBS 0970 is a robust result. As such, with a redshift of z=0.124, RBS 0970 is a relatively distant, spectrally hard, VHE emitting BL Lac object. AGN with these characteristics are ideal for studying the intensity of the extragalactic background light (EBL; \citet{coppi}) and the strength of the intergalactic magnetic field (IGMF; \citet{neronov3}). As such, besides better understanding the VHE properties of RBS 0970, ground-based observations with IACTs will also allow us to use RBS 0970's spectrum to study the EBL and determine the suitability of RBS 0970 for studying the IGMF.

\section{CONCLUSIONS}
With 5.3 years of \textit{Fermi}-LAT data, RBS 0970 has been found to be a source of VHE $\gamma$-ray photons. With 3 \textsc{ultraclean} $E_{\gamma}>100$ GeV photon events within 0.1\ensuremath{^{\circ}} of RBS 0970, an unbinned likelihood analysis revealed the significance of this discovery to be at the $6.5\sigma$ confidence level. The 5.3 year integrated $E_{\gamma}>100$ GeV flux was found to be $(2.47 \pm 1.26) \times 10^{-11}  \text{ photons cm}^{-2} \text{s}^{-1}$. 

An indepth analysis of the $0.1-100$ GeV flux from RBS 0970 during a 28 day window centered on the detection of the VHE photons revealed that the emission of the 114 and 117 GeV photons coincided with a hardening of the $\gamma$-ray spectrum when compared to the 5.3 year average. However, the same analysis did not find any significant $\gamma$-ray emission from RBS 0970 during the emission of the 273 GeV event. This non-detection is hard to accommodate in a pure SSC model description of $\gamma$-ray emission from RBS 0970. 

The detection of 3 \textsc{ultraclean} events within 0.1\ensuremath{^{\circ}}, coupled with the results of the $100-300$ GeV unbinned likelihood analysis suggest the VHE detection of RBS 0970 is a robust result. As such, RBS 0970 is a promising target for follow-up observations with IACTs. Such observations are highly recommended.

\section*{Acknowledgments}

I thank MAB for her helpful discussions and insights which have been invaluable to this paper. I also thank PMC for careful proof reading of this work along with the referee JMcE for her helpful comments and suggestions. This work was undertaken with the financial support of Durham University. This work has made use of public \textit{Fermi} data obtained from the High Energy Astrophysics Science Archive Research Center (HEASARC), provided by NASA’s Goddard Space Flight Center. This work has also made use of the NASA/IPAC Extragalactic Database (NED), which is operated by the Jet Propulsion Laboratory, Caltech, under contract with the National Aeronautics and Space Administration.


\begin{thebibliography}{99}
\bibitem[\protect\citeauthoryear{Abdo et al.}{2009}]{abdo} Abdo, A., et al. 2009, ApJ, 707, 1310
\bibitem[\protect\citeauthoryear{Abdo et al.}{2010}]{1agn} Abdo, A., et al. 2010, ApJS, 715, 429
\bibitem[\protect\citeauthoryear{Abramowski et al.}{2013}]{abram} Abramowski, A., et al. 2013, A\&A, 559, 136
\bibitem[\protect\citeauthoryear{Acharya et al.}{2013}]{cta} Acharya, B.S., et al. 2013, APh, 43, 3
\bibitem[\protect\citeauthoryear{Ackermann et al.}{2011}]{acker3} Ackermann, M., et al. 2011, ApJ, 743, 171
\bibitem[\protect\citeauthoryear{Ackermann et al.}{2012a}]{acker1} Ackermann, M., et al. 2012a, ApJS, 201, 4
\bibitem[\protect\citeauthoryear{Ackermann et al.}{2012b}]{lat2} Ackermann, M., et al. 2012b, ApJS, 203, 4
\bibitem[\protect\citeauthoryear{Ackermann et al.}{2013}]{hecat} Ackermann, M., et al. 2013, ApJS, 209, 13
\bibitem[\protect\citeauthoryear{Aliu et al.}{2012}]{aliu} Aliu, E., et al. 2012, ApJ, 750, 94
\bibitem[\protect\citeauthoryear{Atwood et al.}{2009}]{lat} Atwood, W.B., et al. 2009, ApJ, 697, 1071
\bibitem[\protect\citeauthoryear{Brown}{2006}]{mephd} Brown, A.M., PhD thesis, University of Durham, U.K., 2006
\bibitem[\protect\citeauthoryear{Brown \& Adams}{2011}]{mengc} Brown, A.M. \& Adams, J., 2011, MNRAS, 413, 2785
\bibitem[\protect\citeauthoryear{Brown \& Adams}{2012}]{mepicA} Brown, A.M. \& Adams, J., 2012, MNRAS, 421, 2303
\bibitem[\protect\citeauthoryear{Brown}{2013}]{mepks} Brown, A.M., 2013, MNRAS, 431, 824
\bibitem[\protect\citeauthoryear{Coppi \& Aharonian}{1998}]{coppi} Coppi, P.S. \& Aharonian, F., 1998, 19th Texas Symposium of Relativistic Astrophysics and Cosmology, ed. J. Paul, T. Montemerle \& E. Aubourg
\bibitem[\protect\citeauthoryear{Dickinson \& Farnier}{2013}]{dickinson} Dickinson, H. \& Farnier, C., 2013, A\&A, 552, 134
\bibitem[\protect\citeauthoryear{Elvis et al.}{1992}]{einstein} Elvis, M., Plummer, D., Schachter, J \& Fabbiano,G. 1992, ApJS, 80, 257
\bibitem[\protect\citeauthoryear{Giommi et al.}{2005}]{beppo} Giommi, P., Piranomonte, S., Perri, M. \& Padovani, P, 2005, A\&A, 434, 385  
\bibitem[\protect\citeauthoryear{Holder et al.}{2008}]{veritas} Holder, J., et al. 2008, AIP Conf. Proc. 1085, High Energy Gamma-Ray Astronomy, ed. F.A. Aharonian, W. Hoffman, F.M. Rieger, Vol. 657 (AIP)
\bibitem[\protect\citeauthoryear{Kataoka et al.}{2010}]{kat} Kataoka, J., et al. 2010, ApJ, 715, 554
\bibitem[\protect\citeauthoryear{Krawczynski et al.}{2004}]{kraw} Krawczynski, H. et al. 2004, ApJ, 601, 151 
\bibitem[\protect\citeauthoryear{Macias et al.}{2012}]{oscar} Macias-Ramirez, O., Gordon, C, Brown, A.M. \& Adams, J. 2012, PhRvD, 86, 6004
\bibitem[\protect\citeauthoryear{Mattox et al.}{1996}]{mattox2} Mattox J.R., et al. 1996, ApJ, 461, 396
\bibitem[\protect\citeauthoryear{Neronov et al.}{2010}]{neronov} Neronov, A., Semikoz, D., \& Vovk, Ie., 2010, A\&A, 519, 6
\bibitem[\protect\citeauthoryear{Neronov et al.}{2012}]{neronov2} Neronov, A., Semikoz, D., Taylor, A.M., \& Vovk, Ie., 2012, arXiv: 1207.1962
\bibitem[\protect\citeauthoryear{Neronov \& Semikoz}{2009}]{neronov3} Neronov, A. \& Semikoz, D., 2010, Phys. Rev. D, 80, 123012
\bibitem[\protect\citeauthoryear{Nolan et al.}{2012}]{nolan} Nolan, P.L., et al. 2012, ApJS, 199, 31
\bibitem[\protect\citeauthoryear{Ong \& Fortin}{2010}]{ong} Ong, R. \& Fortin, P., ATel 2272
\bibitem[\protect\citeauthoryear{Padovani \& Giommi}{1995}]{gommi} Padovani, P. \& Giommi, P., 1995, MNRAS, 277, 1477
\bibitem[\protect\citeauthoryear{Plotkin et al.}{2008}]{sdss} Plotkin, R.M., Anderson, S.F., Hall, P.B., Margon, R., Voges, W., Schneider, D.P., Stinson, G. \& York,D.G, 2008, AJ, 135, 2453P
\bibitem[\protect\citeauthoryear{Ritz}{2007}]{ritz} Ritz, S., 2007, Overview of GLAST Mission and Opportunities, Vol. 921 (AIP), 3
\bibitem[\protect\citeauthoryear{Tanaka et al.}{2013}]{tanaka} Tanaka, Y. T, et al. 2013, ApJ, 777, 18
\bibitem[\protect\citeauthoryear{Voges et al.}{1999}]{rosat} Voges, W., 1999, A\&A, 349, 389
\end{thebibliography}
\end{document}